\documentclass{aa}
\usepackage{txfonts}
\usepackage{natbib}
\bibpunct{(}{)}{;}{a}{}{,}
\usepackage{graphicx}
\newcommand{\eqref}[1]{Eq.~(\ref{#1})\relax}
\newcommand{\figref}[1]{Fig.~\ref{#1}\relax}
\usepackage{epstopdf}
\begin{document}

\title{TRACE observations of driven loop oscillations}
\author{Istvan Ballai\inst{1} \and David B. Jess\inst{2} \and Mark Douglas\inst{1}}
\institute{Solar Physics and Space Plasma Research Centre (SP$^2$RC), Department of
Applied Mathematics, The University of Sheffield, Sheffield,
UK, S3 7RH, email: {\tt\{i.ballai;mark.douglas\}@sheffield.ac.uk}
\and Astrophysics Research Centre, School of Mathematics and
Physics, Queen's University Belfast, Belfast, BT7 1NN, Northern Ireland,
UK, email: {\tt d.jess@qub.ac.uk}}
\date{Received dd mmm yyyy / Accepted dd mmm yyyy}

\abstract{} {On 13 June 1998, the TRACE satellite was fortuitously
well placed to observe the effects of a flare-induced EIT wave in
the corona, and its subsequent interaction with coronal magnetic
loops. In this study, we use these TRACE observations to corroborate
previous theoretical work, which determined the response of a
coronal loop to a harmonic driver in the context of ideal
magnetohydrodynamics, as well as estimate the magnetic field
strength and the degree of longitudinal inhomogeneity.} {Loop edges
are tracked, both spatially and temporally, using wavelet modulus
maxima algorithms, with corresponding loop displacements from its
quiescent state analysed by fitting scaled sinusoidal functions. The
physical parameters of the coronal loop are subsequently determined
using seismological techniques.} {The studied coronal loop is found
to oscillate with two distinct periods, $501{\pm}5$~s and
$274{\pm}7$~s, which could be interpreted as belonging to the
fundamental and first harmonic, or could reflect the stage of an
overdriven loop. Additional scenarios for explaining the two periods
are listed, each resulting in a different value of the magnetic
field and the intrinsic and sub-resolution properties of the coronal
loop. When assuming the periods belong to the fundamental kink mode
and its first harmonic, we obtain a magnetic field strength inside
the oscillating coronal loop as $2.0{\pm}0.7$~G. In contrast,
interpreting the oscillations as a combination of the loop's natural
kink frequency and a harmonious EIT wave provides a magnetic field
strength of $5.8{\pm}1.5$~G. Using the ratio of the two periods, we
find that the gravitational scale height in the loop is $73{\pm}3$
Mm.} {We show that the observation of two distinct periods in a
coronal loop does not necessarily lead to a unique conclusion.
Multiple plausible scenarios exist, suggesting that both the derived
strength of the magnetic field and the sub-resolution properties of
the coronal loop depend entirely on which interpretation is chosen.
The interpretation of the observations in terms of a combination of
the natural kink mode of the coronal loop, driven by a harmonic EIT
wave seems to result in values of the magnetic field consistent with
previous findings. Other interpretations which are realistic, such
as fundamental mode/first harmonic and the oscillations of two
sub-resolution threads, result in magnetic field strengths that are
below the average values found before.}


\maketitle
\section{Introduction}

Sudden energy releases in the solar atmosphere are known to generate
large-scale global waves propagating over long distances. The energy
stored in these waves can be released by traditional dissipative
mechanisms, but might also be transferred to magnetic structures
that come into contact with global waves. In this context, EIT waves
generated by coronal mass ejections (CMEs) and/or flares could
interact with coronal loops, resulting in the generation of kink
modes, i.e. oscillations that experience periodic movement about the
loop's symmetry axis. These generated loop oscillations have been
used as a basic ingredient in one of solar physics most dynamically
expanding fields, namely {\it coronal seismology}. Depending on the
orientation of the studied wave, coronal seismology can be further
divided into local
\citep{roberts1984,nakariakov1999,banerjee2007,verth2007} and global
seismology \citep{ballai2005,ballai2007}. Although local and global
seismology appear to be two different aspects of coronal physics, in
reality they are closely related.

Coronal seismology is based on theoretical relations (called
dispersion relations) linking plasma parameters, such as the plasma
density, to wave parameters, such as the wave frequency, in a
precise way. In general, plasma parameters are determined from wave
parameters, which themselves are determined observationally. The
dispersion relations for many simple (and some quite complicated)
plasma structures under the assumptions of ideal
magnetohydrodynamics (MHD) are well known; they were derived long
before accurate EUV observations were available
\citep{edwinroberts1983,roberts1984} using simplified models within
the framework of ideal and linear MHD. Although the accurate
interpretation of many observations can be difficult because of the
insufficient spatial and temporal resolutions of present satellites,
considerable amounts of information about the state of the plasma,
and the structure and magnitude of the coronal magnetic field, can
still be obtained.

Most of the previous studies of waves and oscillations in coronal
loops have focussed on the (local) oscillation itself, without
considering the more global cause of the oscillation. One of the
first studies where the nature of a local loop oscillation was
investigated in terms of the type of driver (global EIT waves) was
the study by \cite{ballai2008}. These authors analyzed the pattern
of possible oscillations (treated as an initial value problem) that
can be recovered in a simplified loop model under the influence of,
e.g. harmonic drivers interacting with loops. Their main conclusion,
which is used in the present study, was that for a certain range of
driver periods, the loop supports oscillations whose period is a
superposition of the driver's period and the natural period of the
loop. A similar analysis of a 2D loop was carried out by
\cite{selwa10}.

The aim of this paper is to investigate, using TRACE EUV
observations, the characteristics of a coronal loop that oscillates
under the influence of an external driver. The special circumstances
of this event reside in the possibility of connecting the loop
oscillation to a global EIT wave, presenting the phenomena of loop
oscillations in a much wider context.

\section{Observations and data analysis}

In contrast to the SOHO/EIT instrument, the TRACE/EUV imager has
good temporal and spatial resolution, suitable for observing waves
in the MHD domain. However, the field-of-view of TRACE is merely 8.5
minutes of arc, compared to SOHO/EIT's full disk images of 45
minutes of arc. Thus, obtaining TRACE data close to the source of an
EIT wave requires a small measure of good fortune; one such instance
occurred on 13 June 1998. On this day, a coronal wave was initiated
just south of the field of view of TRACE at approximately 15:23 UT
by a GOES C2.9 flare. The maximum flare intensity occurred between
the eastern part of the main negative polarity and the western part
of the main positive polarity regions of AR~8237 (S25W04 in the
solar heliographic coordinate system) at 15:33~UT, where a filament
was also observed in H$\alpha$. Furthermore, this event was
associated with a halo CME observed in white light by LASCO C2 with
a leading edge moving at 190 kms$^{-1}$ \citep{delannee2000}. A
limited study of this event has already been carried out by
\cite{wills-davey1999}, with the authors interpreting the
disturbance as a fast magnetoacoustic wave.

\begin{figure}
\centering
\includegraphics[width=\columnwidth]{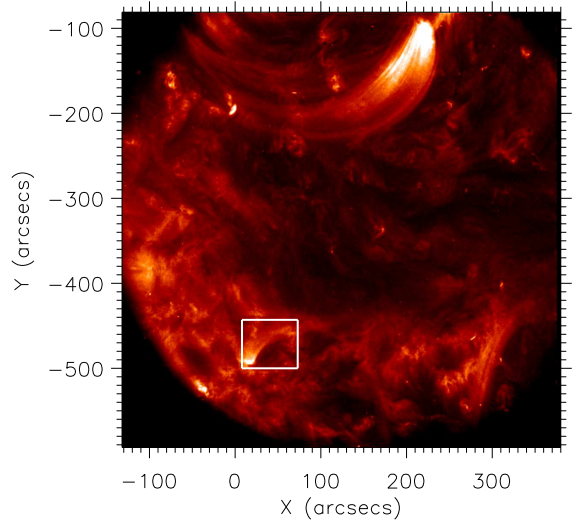}
\caption[TRACE field of view]{TRACE field-of-view at 171~{\AA}
immediately prior to the arrival of the EIT wave front.
The white box indicates the coronal loop under investigation
in the present study. Axes are in solar heliocentric arcseconds.}
\label{fig1}
\end{figure}

\begin{figure*}
\centering
\includegraphics[width=16cm]{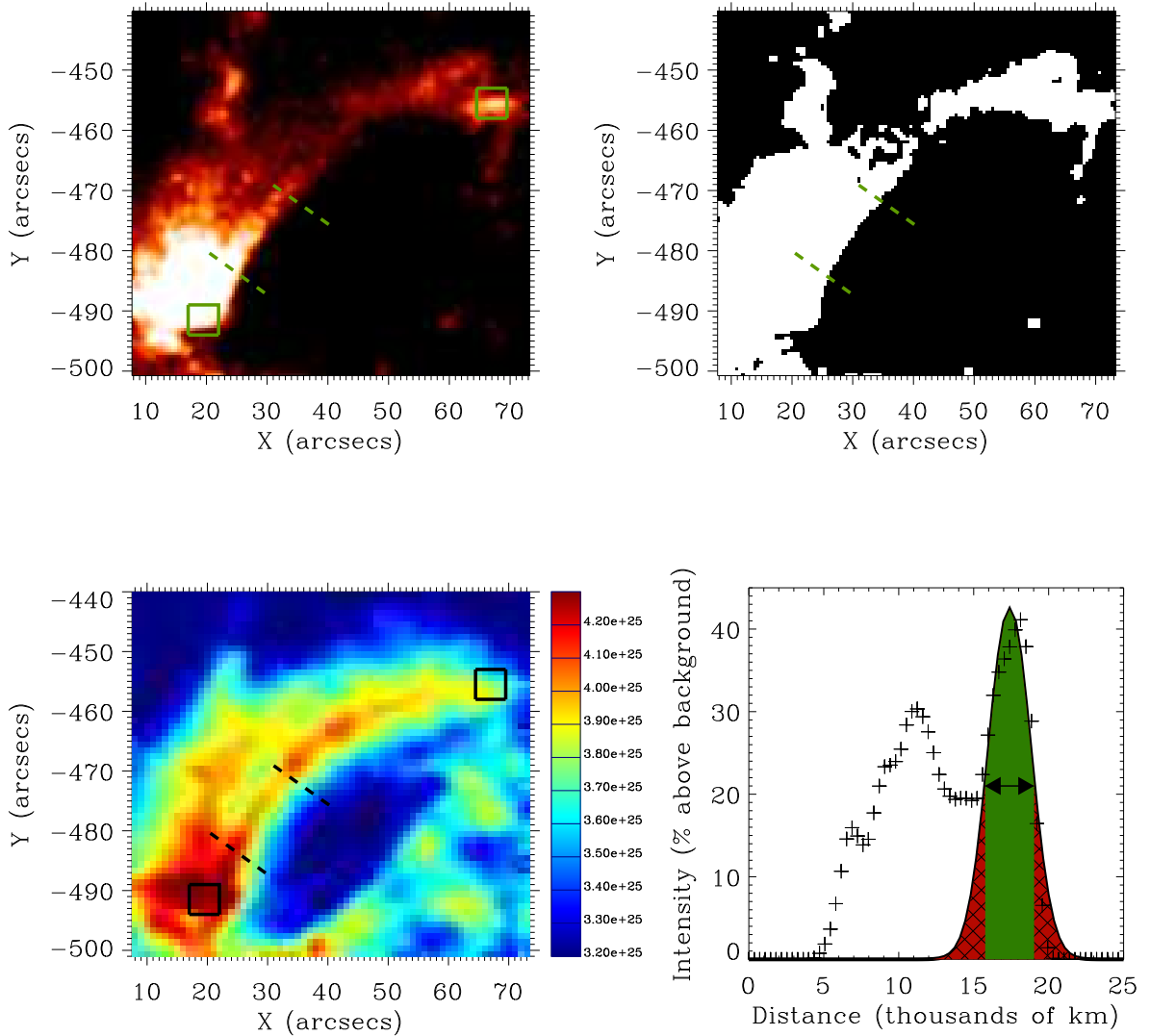}
\caption[Small TRACE field of view]{Zoomed images of the TRACE
field-of-view contained within the white box in \figref{fig1}. The
upper-left image displays a true intensity image, while the
upper-right panel reveals the corresponding loop-edge binary map
used in the tracking algorithm of \citet{Jess08}. Locations of the
loop footpoints are highlighted with green boxes in the left panel,
giving an approximate loop length, $L = 72.9{\pm}2.8$~Mm. An
11000{\,}km segment, incorporating a brightly emitting section of
the loop edge, which is used in the search for higher harmonics is
contained within the green dashed lines plotted in the upper panels.
The lower-left panel displays an emission measure map, derived from
near-simultaneous 171{\,}{\AA} and 195{\,}{\AA} TRACE images, with
the corresponding color scale in units of cm$^{-5}$. Black boxes and
dashed lines represent characteristics identical to those described
above for the upper-left panel. The loop arcade intensity ($+$),
averaged between the green dashed lines in the upper-left panel, is
plotted in the lower-right panel as a percentage above the quiescent
background. A Gaussian fit to the coronal loop under investigation
is displayed using a solid black line. Green shading represents loop
intensities greater than the fitted half-maximum, while
cross-hatched red shading indicates regions of the loop structure
with intensities below the half-maximum. Arrows mark the full-width
at half-maximum, chosen to represent an estimate of the loop depth,
found here to be $\approx$8~pixels, or $\approx$3000{\,}km.}
\label{fig2}
\end{figure*}

TRACE images in the 171~{\AA} and 195~{\AA} bandpasses acquired
during 15:00 -- 19:00~UT show the progress of the wave (activity
over an extended area of the field-of-view), and from the SOHO/EIT
instrument a concomitant EIT wave is visible
\citep{wills-davey1999}. The TRACE data also reveal how the EIT wave
front causes the positioning of some coronal loops to be disturbed.
The particular loop under investigation is situated in the
south-eastern part of \figref{fig1}, with a zoom into the region of
interest in \figref{fig2}.

Using the standard calibration/correction packages available in the
{\it SolarSoft} library, the TRACE data was processed in the usual
way: spikes from cosmic ray hits were removed, data frames were
normalized to a constant exposure time, and the image sequence was
de-rotated. The TRACE satellite entered the Southern Atlantic
Anomaly at approximately 15:50 UT, and images up to this point
clearly display a propagating wave front.

\subsection{Temporal analysis of loop edge}
\label{WMM} The characteristics of a global EIT wave were studied in
detail by \citet{ballai2005}, who demonstrated that the global
disturbance is a wave with an average periodicity of approximately
400~s. To accurately probe the effects of this wave front on more
rigidly fixed coronal loop structures, the loop edge was tracked
spatially and temporally using a wavelet modulus maxima technique
described by \citet{Jess08}. Since the 195~{\AA} response function
is somewhat lower than that of the 171~{\AA} passband, longer
exposure times are required to provide the same signal as the
171~{\AA} channel (a 41{\,}s average exposure time at 195~{\AA},
versus 24{\,}s at 171~{\AA} images). As a result, rapidly moving
structures may be blurred in the 195~{\AA} passband because of their
substantially longer exposure times. Furthermore, the 171~{\AA}
images are at full resolution (1 pixel = 0.499$''$), whereas the
195~{\AA} data are $2{\times}2$ binned, resulting in a pixel size of
0.998$''$. Through use of blind iterative deconvolution techniques,
\citet{Gol99} and \citet{Lin01} determined the absolute spatial
resolution of the TRACE instrument to be 1.25$''$. Since the
171~{\AA} data is not degraded by spatial binning, it enables us to
monitor small-scale disturbances more clearly in coronal loop
structures. As a result, only images taken in the 171~{\AA} bandpass
could be tracked with an average cadence of 90~s. First, a segment
of the loop edge, between the loop footpoint and apex, was passed
through a Laplacian filter to sharpen the images and make the edge
of the structure more pronounced. Next, a lower intensity threshold
was applied to the data to remove contributions from the underlying
quiet Sun. A threshold of the background mean flux plus 4.5 $\sigma$
was used, which produces the image displayed in the upper-left panel
of \figref{fig2}. To emphasize feature edges and remove shallow
intensity gradients, we used a binary format for the feature
mapping. All pixels of values above the lower flux intensity
threshold defined above were assigned a value of `1'. Those pixels
which lie below the threshold are assigned a value of `0', producing
the binary map shown in the upper-right panel of \figref{fig2}. To
assist the analysis, each image was rotated by 45$^{\circ}$
clockwise until the southern edge of the loop structure was parallel
with the x-axis. This caused loop displacements to be confined to
the direction of the y-axis, as a function of the loop length
(x-axis). TRACE images acquired between the initiation of
oscillatory behaviour at 15:29~UT, and 15:50~UT when the instrument
entered the Southern Atlantic Anomoly, were binary mapped for edge
detection and tracking using the methodology described by
\citet{Jess08}.

\begin{figure*}
\centering
\includegraphics[width=16cm]{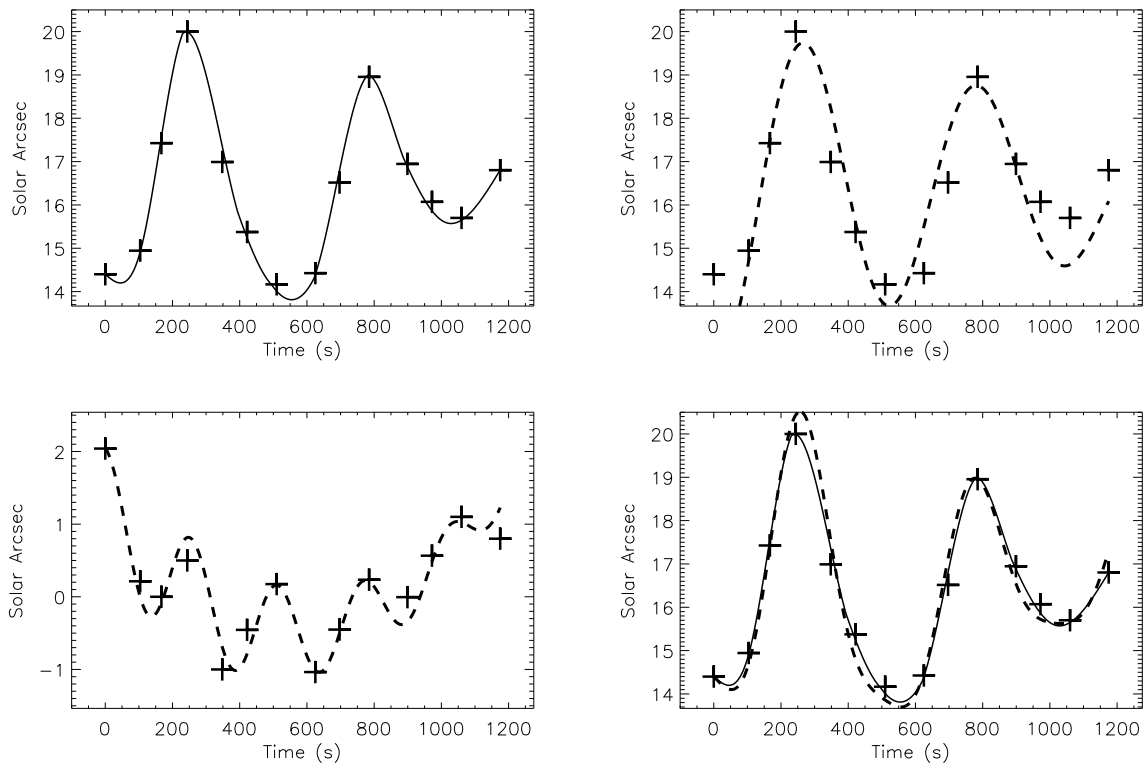}
\caption[Sine fitting analysis of loop segment]{Top Left: A typical
transverse displacement of one pixel along the loop segment ($+$),
joined together using a spline fit to assist visualisation of the
motion path (solid line). Top Right: The original transverse displacement ($+$),
overplotted with a 501{\,}s period sinusoidal signal
(dashed line). Bottom Left: The residual (501{\,}s period sinusoid subtracted
from the original motion path) transverse displacement ($+$), overplotted
with a 274{\,}s period sinusoidal signal (dashed line). Bottom Right: The
original data points ($+$) and spline fit (solid line) shown in the upper-left
panel, overplotted with the combined 501/274{\,}s period sinusoidal fits. These
combined fits, with respect to the original time series, result in
Kolmogorov-Smirnov and cross-correlation statistics of 92\% and 98\%,
respectively, indicating a high degree of accuracy.}
\label{fig3}
\end{figure*}

Utilizing the resulting loop-edge positions, the displacement of the
loop segment as a function of time from its quiescent state was
established for 30 pixels (${\approx}11000$~km; see dashed lines in
\figref{fig2}) along the loop edge between the footpoint and apex.
Owing to the minimal shifts in the loop equilibrium position with
time, and the need to minimise the number of free fitting
parameters, no long-term trends were subtracted from the original
time series. The loop-edge displacement curve was then fitted using
sinusoids of varying frequency and amplitude. The best fit sinusoid
was then subtracted from the original data to produce a residual
time series. The residual data, again, was modelled with the
best-fit sinusoid function. To determine the success of the fitting
process, Kolmogorov-Smirnov and cross-correlation statistics
\citep{Press92, Christian98, Jess07a, Jess07b} were determined for
each stage of the sinusoid fitting process, as well as for the
combined fit with respect to the original time series. An example of
this process is displayed in \figref{fig3}, where Kolmogorov-Smirnov
statistics for the 501{\,}s, 274{\,}s, and combined fits are 88\%,
91\%, and 92\%, respectively. Cross-correlation statistics for the
same fitting functions are 90\%, 74\%, and 98\%, respectively,
indicating that these sinusoid functions recreate the observed data
with a high degree of precision.

To validate this result, the process of loop isolation, tracking,
and sinusoid-fitting was repeated for image sequences that had been
spatially degraded using $2{\times}2$, $3{\times}3$, $5{\times}5$,
and $7{\times}7$ binning intervals. While spatial binning has no
effect on a robust periodic signature, it will help us to reduce any
artefacts present in the data, such as those caused by detector
noise and/or cosmic ray hits. These independent analyses yielded
consistent results, for which periodicities overlapped with those
established above, implying that the oscillation occurs over a large
spatial scale, and is unaffected by varying smoothing intervals. As
described in \citet{vandoorsselaere2007}, repeating this procedure
for, and averaging over, many loop positions and binning intervals
($30$ loop positions ${\times}$ $5$ binning values $= 150$
independent measurements in this instance) allows the associated
errors to be drastically reduced. Following a similar methodology,
we derive best-fit periodicities of $501{\pm}5$~s and $274{\pm}7$~s,
both of which are above the limiting Nyquist period ($\approx$180~s)
of the dataset.

\subsection{Properties of the coronal loop}
The fact that two different periods have been recovered from the
same structure allows us to uncover properties of the loop that
cannot be directly measured. We use a simple semicircular model for
the loop, where the loop length, $L$, is given by $\pi D/2$, where
$D$ is the distance between footpoints. As the TRACE field of view
is nearly at the solar disk centre, and the loop does not appear to
be tilted at a significant angle, this model is likely to be
accurate. The distance $D$ is measured to be $128{\pm}5$ TRACE
pixels, corresponding to an absolute diameter of $D =
46.4{\pm}1.8$~Mm. Thus, we are dealing with a relatively short loop,
with a corresponding estimated loop length of $L = 72.9{\pm}2.8$~Mm.

The electron number density of the loop (and its environment) can be
estimated using a differential emission measure technique by
comparing the intensities of the loop detected at different
wavelengths. This filter-ratio method is based on the assumption
that a structure seen in both 171{\,}{\AA} and 195{\,}{\AA}
bandpasses will be isothermal, with an associated temperature
falling between the two peak temperatures of the TRACE filter
response functions \citep[]{Asch00}. Using near-simultaneous
171{\,}{\AA} and 195{\,}{\AA} TRACE images, and following the
methodology described in \citet{aschwanden}, we calculate the
emission measure, $EM$, as $EM = (4.0{\pm}0.2) \times
10^{25}$~cm$^{-5}$, averaged over the 30-pixel loop segment used in
Section~\ref{WMM} (lower-left panel of \figref{fig2}). This value
represents the emission integrated over the entire line-of-sight.
The relationship between emission measure and electron number
density, $n_e$, in a source plasma structure of depth, $d$, is given
by
\begin{equation}
n_e(x,y) = \sqrt{\frac{EM(x,y)}{d}} \ ,
\end{equation}
where $x$ and $y$ are the spatial coordinates \citep{aschwanden}.
Assuming a cylindrical loop structure, and using the full-width at
half maximum to estimate the loop cross-section, we estimate the
loop depth to be $8{\pm}1$ pixels, or $(29.2{\pm}3.6) \times
10^7$~cm (lower-right panel of \figref{fig2}). This allows us to
calculate the electron number density in the loop as $n_e =
(3.7{\pm}2.3) \times 10^8$~cm$^{-3}$. Using a quiescent plasma
electron number density (the region outside the loop) of $1\pm 0.5
\times 10^8$~cm$^{-3}$ in line with previous estimates (see, e.g.
Fludra et al. 1999, Ugarte-Urra et al. 2002), we can determine the
loop filling factor to be $3.7{\pm}0.4$. For brightly emitting
coronal loops, it can be assumed that all of the emission measure
comes directly from the loop itself. However, for relatively dim
loops, similar to the one under investigation here, a significant
portion of the emission will come from plasma in the immediate
vicinity of the loop. As a result, the derived electron number
densities will represent the upper boundaries of actual values.

\subsection{Application of coronal seismology}
Coronal seismology, initiated after high-resolution observations of
waves and oscillations in coronal structures became available, can
provide relatively accurate estimates of quantities that cannot be
measured directly or are below the instrument's resolution limit.
Using a combination of observations and theoretical models, answers
to the problems of the magnitudes and structuring of coronal
magnetic fields, density scale-heights, etc., have since been
recovered
\citep{nakariakov1999,andries2005,andries2009,VerthErdelyiJess2008,
RudermanErdelyi2009}.

\begin{figure}
\centering
\includegraphics[width=\columnwidth]{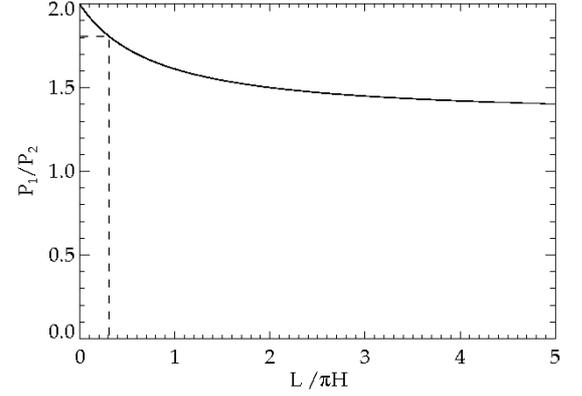}
\caption[The period ratio between the fundamental mode and the first
harmonic] {The period ratio, $P_1/P_2$, corresponding to the
fundamental to first harmonic kink mode frequencies, plotted as a
function of loop length, $L$, and density scale height, $H$. A
$P_1/P_2$ ratio equal to 1.82, combined with a loop length
$L{\approx}73$~Mm, provides a density scale height of
$H{\approx}72$~Mm.} \label{fig4}
\end{figure}

In this context, one of the most promising methods for investigating
the internal structure of coronal loops is based on the $P_1/P_2$
ratio of the fundamental ($P_1$) to first harmonic ($P_2$) modes of
an oscillation. In a homogeneous, non-stratified atmosphere the
period ratio is 2. However, longitudinal variations in the
properties of the plasma (i.e. variations along the loop) can make
this ratio to deviate either below or above 2. It is believed that
magnetic effects produce a ratio that is greater than 2 because of
the magnetic divergence caused by changes in the cross-sectional
area of the loop \citep{DeMoortelBrady2007,oshea2007}. In this case,
the magnetic effect on the period of oscillation dominates any other
effects, such as density stratification. In contrast, if the ratio
of the two periods is less than 2, then the change is attributed to
density stratification effects along the coronal loop. Several
examples have been published where the $P_1/P_2$ ratio is within the
interval $1.5-1.9$
\citep{mcewan2006,vandoorsselaere2007,ruderman2008,VerthErdelyi2008,VerthErdelyiJess2008}.
\cite{andries2005} show how a one-to-one relationship exists between
the $P_1/P_2$ ratio and the density scale-height. The dependence of
the period ratio on $L/{\pi}H$, where $H$ is the density
scale-height, is shown in \figref{fig4}.

According to \cite{andries2005}, the dependence of the period ratio
on the density stratification does not depend on other loop
parameters or the magnetic field strength. As a result, the same
dependence can be demonstrated for a variety of density profiles
generated inside the same coronal loop. For the coronal loop under
current investigation, we find that $P_1/P_2 = 1.82{\pm}0.02$,
corresponding to a density scale height $H = 73{\pm}3$~Mm. This
value is consistent with those found by \citet{vandoorsselaere2007}.

Assuming that the loop is in hydrostatic equilibrium, the density
scale height is directly proportional to the temperature, $T$,
according to the equation
\begin{equation}
H = 47\left(\frac{T}{MK}\right) \ ,(Mm).
\label{height}
\end{equation}
Thus, for our present analysis we obtain a loop temperature of $T =
1.5{\pm}0.6$~MK, which is within TRACE's temperature response for
the 171~{\AA} channel, and towards the central portion of the
195~{\AA} channel's temperature response curve
\citep{aschwanden,phillips}. Nevertheless, the visibility of the
coronal loop in both the 171~{\AA} and 195~{\AA} bandpasses, in
addition to a derived loop temperature within the TRACE instrument's
temperature response function, suggests that the loop under
investigation is in hydrostatic equilibrium \citep{andries2005}.

On the basis of the period of the fundamental mode (501~s),
following the methods of \citet{verwichte2004}, it is possible to
calculate the kink speed, $c_K$, of the wave to be $c_K =
2L/P_{\mathrm{fundamental}} = 291{\pm}8.2$~kms$^{-1}$. The kink
speed can then be related to the magnetic field strength, $B$, by
the equation
\begin{equation}
B=c_K\left[\frac{\mu(\rho_i+\rho_e)}{2}\right]^{1/2},
\label{eq:kinkmag}
\end{equation}
where $\rho_i$ and $\rho_e$ are the internal and external loop
densities, respectively, $\mu$ is the permeability of free space,
and the magnetic field strength inside and outside the loop is
assumed to be identical. Thus, we estimate the magnetic field
strength {\it inside} the coronal loop to be $2.0{\pm}0.7$~G, which
is towards the lower-end of statistical studies compiled by
\citet{Nak01} and \citet{asch02}.

\subsection{Comparison with other theories}
The existence of two distinct periods in the loop under
investigation has been explained and investigated assuming that the
two periods belong to the fundamental and first harmonic kink
oscillations of the loop. However, other viable explanations are
also possible. For our discussion, we first recall the results of
\cite{ballai2008}. Using a simple coronal loop model, these authors
investigated the interaction between an EIT wave and a coronal loop
assuming an equilibrium of forces. Their results show that the
periods recovered in a coronal loop are always a combination of the
period of the driver (here EIT waves) and the natural period of the
loop. The same authors showed that the displacement of a coronal
loop (here denoted by $Q(z,t)$) under the influence of a driver,
$F(z,t)$, is given by
\begin{equation}
\frac{\partial^2Q}{\partial t^2}-c_K^2\frac{\partial^2Q}{\partial z^2}+\omega_c^2Q=F \ ,
\label{eq:3.1}
\end{equation}
where, as before, $c_K$ is the kink speed (propagation speed of
disturbances in the loop) and $\omega_c$ is the cut-off frequency of
kink-mode oscillations. These quantities are simply given by
\[
c_K=\sqrt{\frac{\rho_iv_{Ai}^2+\rho_ev_{Ae}^2}{\rho_i+\rho_e}} \ ,
\quad \omega_c=\frac{g(\rho_i-\rho_e)}{2c_K(\rho_i+\rho_e)} \ ,
\]
where $v_{Ai}$ and $v_{Ae}$ are the Alfv\'en speeds inside and
outside the loop. Equation~\ref{eq:3.1} is an inhomogeneous
Klein-Gordon equation that was solved assuming that the foot-points
of the loop are fixed (line-tying condition) and initially at rest.

The event occurring on 13 June 1998 was also studied in detail by
\cite{wills-davey1999}, who showed that the blast wave intersects
the loop of interest nearly perpendicularly, meaning we can model
the EIT wave as a harmonic driver whose temporal and spatial
dependence is given by \cite{ballai2008} to be
\begin{equation}
F(z,t)\propto
K\left[\delta(z-z_0)-\delta(z-L+z_0)\right]e^{i\omega_{EIT}t} \ ,
\label{eq:3.2}
\end{equation}
where $\delta(z)$ is the Dirac-delta function, and the constant $K$
depends on the density of the plasma and energy of the EIT wave,
both considered to be constant. In Equation~\ref{eq:3.2},
$\omega_{EIT}$ is the frequency of the EIT wave, which is assumed to
be a harmonic signal. In this study, we are concerned with temporal
changes only, which is why the exact form of the driver is not
explicitly given. The form of the driver given by
Equation~\ref{eq:3.2} reveals that the interaction between the loop
and the EIT wave occurs in two points (at $z=z_0$ and $z=L-z_0$)
simultaneously, and is placed symmetrically with respect to the
end-points of the loop at $z=0$ and $z=L$.

Equation~\ref{eq:3.1}, where $F(z,t)$ is defined by
Equation~\ref{eq:3.2}, was solved by \cite{ballai2008}, with the
temporal dependence of the displacement, $Q(z,t)$, found to be
\begin{equation}
Q(z,t)\propto \sum_{n=1}^{\infty}
\sin\left[\left(\frac{\omega_n+\omega_{EIT}}{2}\right)t\right]
\sin\left[\left(\frac{\omega_n-\omega_{EIT}}{2}\right) t\right] \ ,
\label{eq:3.3}
\end{equation}
where $\omega_n^2=\omega_c^2+n^2\pi^2c_K^2/L^2$ ($n\ge 1$) are the
natural frequencies of the loop.

According to the results of \cite{ballai2008} (their Figure 3), the
dominant (and strongest) signal in the loop is produced by the
driver. Bearing in mind the previously presented model, we identify
the two measured periods as the two temporal dependencies of the
solution given by Equation~\ref{eq:3.3}. Since it is visually
obvious that the loop oscillates at its fundamental frequency,
$\omega_n$ in Equation~\ref{eq:3.3}, it is replaced by $\omega_1$
corresponding to the frequency of the fundamental mode. That means
that we need to solve the system
\begin{equation}
\omega_1+\omega_{EIT}=\frac{4\pi}{T_2}, \quad
\omega_1-\omega_{EIT}=\frac{4\pi}{T_1}, \label{eq:3.5}
\end{equation}
where $T_1=501$~s and $T_2=274$~s. It is easy to show that the
period of the fundamental mode is $177.1\pm 3.5$ s and the period of the
driving EIT wave is $604.7\pm 26.8$ s, much higher than the period derived by
\cite{ballai2005}. However, we should keep in mind that the result derived
by \cite{ballai2005} refers to an average period derived from
within a large field-of-view.

Once the period of the fundamental mode is derived, it is
straightforward to estimate the magnetic field strength inside the
loop. We, again, need to use the mathematical finding of
\cite{ballai2008} to derive the magnetic field. According to their
analysis, the frequency of the fundamental mode is given by
\begin{equation}
\omega_1=\sqrt{\omega_c^2+\pi^2c_K^2/L^2}. \label{eq:3.6}
\end{equation}
Using the previously determined values for densities and length of
the loop, together with the derived frequency of the fundamental
mode, we obtain the fundamental kink speed of the loop to be
$823\pm 15$\,km\,s$^{-1}$.
Using the standard definition of the kink speed, and assuming the
same magnetic field strength inside and outside the loop, we can
obtain the magnetic field in the loop to be
\begin{equation}
B=c_K\sqrt{\frac{\mu(\rho_i+\rho_e)}{2}}=5.8\pm 1.8\; G \ .
\label{eq:3.8}
\end{equation}
The values of the magnetic field strength determined using two
different scenarios are rather different, meaning that the
determination of this fundamental quantity depends on the applied
theoretical model used to explain the observed periods.

As we specified earlier the interpretation of two periods observed
in a loop is not unique because different scenarios can occur.
Unfortunately, given the present spatial resolution constraints, it
is impossible to distinguish between these models, meaning that all
findings derived from the observations of loop oscillations should
be treated with care.

Another possible explanation of the two observed periods is
connected to the limits of the spatial resolution of the TRACE
instrument. The two observed periods could belong to two neighboring
thin loops that cannot be resolved by our observations moving in
phase. This possibility was studied using numerical simulations by
\cite{luna2008}, and using theoretical models by \cite{vandoors08},
\cite{robertson10}, and \cite{robertson11}. Assuming that the loops
are identical, we can easily find that the ratio of $d/R$, where $d$
is the distance between the longitudinal axes of the two loops and
$R$ is their radius, is $2.00\pm0.11$. Thus, the physical distance
between the two loops, in units of their radius, is $0.031\pm
0.001$. Furthermore, using the results of \cite{vandoors08}, we can
estimate the magnetic field inside the loop to be $2.6\pm0.4$~G, a
value remarkably consistent with that obtained when interpreting the
two periods as belonging to the fundamental and first harmonic kink
mode of a single coronal loop.

Finally, the fourth possible explanation is related to the intrinsic
properties of the coronal loop where the second harmonic is not a
proper eigenmode but just reflects a slightly more triangular
amplitude profile than sinusoidal. In other words, the second period
is observed as some nonlinear amplitude oscillation caused by a
gradient in either density or magnetic field across the loop motion,
or a coupling with neighbouring loops.

\section{Conclusions}

To carry out coronal seismology it is imperative to know the
relationship between the composition of a plasma structure and the
oscillations supported by the coronal loop. High resolution
observations enable us to accurately measure not only the magnetic
field strength, but also the thermodynamical state of the plasma.

The event studied here occurred on 13 June 1998, and was recorded by
the TRACE instrument in both 171~{\AA} and 195~{\AA} bandpass
filters. A sudden energy release outside the field-of-view generated
a global wave that interacted with local coronal loops. The
resulting loop oscillations were identified as kink modes, i.e.
oscillations that propagate along the coronal loop leading to the
displacement the longitudinal symmetry axis of the tube. Rigorous
analysis of the oscillation revealed the existence of two distinct
periods contained within a relatively short loop. Periodicities of
$501{\pm}5$~s and $274{\pm}7$~s allowed us to draw conclusions about
the magnetic field strength and the longitudinal density structure
inside the loop. The magnetic field established here has a value
towards the lower end of recent statistical studies. However, a 30\%
increase in the magnetic field strength can be achieved if the loop
filling factor is doubled. A major conclusion of our study is that
two observed periodicities in a loop structure might not be a unique
interpretation, often with derived system parameters differing
substantially owing to the independent seismological methods. We
have compared the observational evidence with two different
theoretical interpretations of oscillations. First, we considered
that the two periods are the periods corresponding to the
fundamental and first harmonic. The $P_1/P_2 = 1.82{\pm}0.02$ ratio
of the two periods has allowed us to determine values of
$73{\pm}3$~Mm and $1.5{\pm}0.6$~MK for the density scale height and
loop temperature, respectively. Furthermore, an estimate of the
magnetic field strength inside the coronal loop was determined to be
$2.0{\pm}0.7$~G. All values determined are consistent with previous
studies involving coronal seismology, even though the coronal loop
structure presented here is relatively confined and dim.

A second possible explanation of two periods might be that developed
by \cite{ballai2008}, which states that the two periods might belong
to the driver (here an incident EIT wave assumed to be harmonic) and
the visibly clear fundamental mode kink oscillation. Using the
results of \cite{ballai2008}, we obtained the period of the
independent driver and kink oscillation, with the latter quantity
allowing us to again estimate the magnetic field strength inside the
loop structure. The theory developed by \cite{ballai2008} relies
heavily on the EIT wave being described using wave theory, as
opposed to a propagating magnetic feature. Our observations provide
additional support that wave theory can describe the mechanism
behind EIT waves. As such, we feel the use of EIT waves for coronal
seismology is an accurate and reliable approach. The magnetic field
determined in light of this scenario is almost triple that obtained
utilising the first method. On the basis of the magnetic field
strength alone, and its consistency with previous estimated values,
we deem that the interpretation of the two periods as the
combination of the loop's natural fundamental period and the period
of the harmonic driver, seems more plausible. Furthermore, when the
two periods were assumed to belong to two unresolved coronal threads
we were able to estimate both the magnetic field inside the threads,
in addition to the distance between them. Finally, another possible
explanation of the two periods was linked to the significant density
or magnetic field gradient along the motion path of the oscillating
loop, or an aspect of coupling with neighbouring loop structures.

Our study reveals that the periods of oscillations of a coronal loop
should be interpreted with great care as a multitude of scenarios
are plausible that would each result in different values of, e.g.
magnetic field. Further higher resolution observational evidence
will help us to apply the most likely interpretation of dynamical
processes.

\acknowledgements

IB acknowledges the financial support by NFS Hungary (OTKA, K83133).
DBJ wishes to thank the Science and Technology Facilities Council
for the award of a Post-Doctoral Fellowship. MD acknowledges the
support of STFC. We are grateful to the anonymous referee for
his/her comments and for pointing out the fourth possible
explanation of two periods in a coronal loop.

\vspace{1cm}

\bibliography{aa}
\bibliography{submit}
\bibitem[Andries et~al.(2005)]{andries2005}
Andries, J., Arregui, I., \& Goossens, M., 2005, \apjl, 624, L57
\bibitem[Andries et~al.(2009)]{andries2009}
Andries, J., van Doorsselaere, T., Roberts, B., Verth, G., Verwichte, E., \& Erd{\'e}lyi, R., 2009, \ssr, 149, 3
\bibitem[Aschwanden(2004)]{aschwanden}
Aschwanden, M.~J., 2004, Physics of the Solar Corona, Springer-Verlag, Berlin
\bibitem[Aschwanden et~al.(2002)]{asch02}
Aschwanden, M.~J., de Pontieu, B., Schrijver, C.~J., \& Title, A.~M., 2002, \solphys, 206, 99
\bibitem[Aschwanden et al.(2000)]{Asch00}
Aschwanden, M.~J., Tarbell, T.~D., Nightingale, R.~W., Schrijver, C.~J., Title, A.,
Kankelborg, C.~C., Martens, P., \& Warren, H.~P., 2000, \apj, 535, 1047
\bibitem[Ballai et~al.(2005)]{ballai2005}
Ballai, I., Erd{\'e}lyi, R., \& Pint{\'e}r, B., 2005, \apjl, 633, L145
\bibitem[Ballai(2007)]{ballai2007}
Ballai, I., 2007, \solphys, 246, 177
\bibitem[Ballai et~al.(2008)]{ballai2008}
Ballai, I., Douglas, M., \& Marcu, A., 2008, \aap, 488, 1125
\bibitem[Banerjee et~al.(2007)]{banerjee2007}
Banerjee, D., Erd{\'e}lyi, R., Oliver, R., \& O'Shea, E., 2007, \solphys, 246, 3
\bibitem[Christian et~al.(1998)]{Christian98}
Christian, D.~J., Drake, J.~J., \& Mathioudakis, M., 1998, \aj, 115, 316
\bibitem[De Moortel \& Brady(2007)]{DeMoortelBrady2007}
De Moortel, I., \& Brady, C.~S., 2007, \apj, 664, 1210
\bibitem[Delann{\'e}e(2000)]{delannee2000}
Delann{\'e}e, C., 2000, \apj, 545, 512
\bibitem[Edwin \& Roberts(1983)]{edwinroberts1983}
Edwin, P.~M., \& Roberts, B., 1983, \solphys, 88, 179
\bibitem[Fludra et al.(1999)]{fludra99}
Fludra, A., Del Zanna, G., Alexander, D. \& Bromage, B. J. I., 1999, \jgr, 104, 9709
\bibitem[Golub et al.(1999)]{Gol99}
Golub, L., et al., 1999, Physics of Plasmas, 6, 2205
\bibitem[Jess et~al.(2007a)]{Jess07a}
Jess, D.~B., Andi{\'c}, A., Mathioudakis, M., Bloomfield, D.~S., \& Keenan, F.~P., 2007, \aap, 473, 943
\bibitem[Jess et~al.(2007b)]{Jess07b}
Jess, D.~B., McAteer, R.~T.~J., Mathioudakis, M., Keenan, F.~P., Andic, A., \& Bloomfield, D.~S., 2007, \aap, 476, 971
\bibitem[Jess et~al.(2008)]{Jess08}
Jess, D.~B., Mathioudakis, M., Erd{\'e}lyi, R., Verth, G., McAteer, R.~T.~J., \& Keenan, F.~P., 2008, \apj, 680, 1523
\bibitem[Lin et al.(2001)]{Lin01}
Lin, A.~C., Nightingale, R.~W., \& Tarbell, T.~D., 2001, \solphys, 198, 385
\bibitem[Luna et~al.(2008)]{luna2008}
Luna, M., Terradas, J., Oliver, R., \& Ballester, J.~L., 2008, \apj, 676, 717
\bibitem[McEwan et~al.(2006)]{mcewan2006}
McEwan, M.~P., Donnelly, G.~R., D{\'{\i}}az, A.~J., \& Roberts, B., 2006, \aap, 460, 893
\bibitem[Nakariakov et~al.(1999)]{nakariakov1999}
Nakariakov, V.~M., Ofman, L., Deluca, E.~E., Roberts, B., \& Davila, J.~M., 1999, Science, 285, 862
\bibitem[Nakariakov \& Ofman(2001)]{Nak01}
Nakariakov, V.~M., \& Ofman, L., 2001, \aap, 372, L53
\bibitem[O'Shea et~al.(2007)]{oshea2007}
O'Shea, E., Srivastava, A.~K., Doyle, J.~G., \& Banerjee, D., 2007, \aap, 473, L13
\bibitem[Phillips et~al.(2005)]{phillips}
Phillips, K.~J.~H., Chifor, C., \& Landi, E., 2005, \apj, 626, 1110
\bibitem[Press et~al.(1992)]{Press92}
Press, W.~H., Teukolsky, S.~A., Vetterling, W.~T., \& Flannery, B.~P., 1992, Cambridge: University Press, |c1992, 2nd ed.,
\bibitem[Roberts et~al.(1984)]{roberts1984}
Roberts, B., Edwin, P.~M., \& Benz, A.~O., 1984, \apj, 279, 857
\bibitem[Robertson et~al.(2010)]{robertson10}
Robertson, D., Ruderman, M.S., \& Taroyan, Y., 2010, \aap, 515, 33
\bibitem[Robertson \& Ruderman(2011)]{robertson11}
Robertson, D. \& Ruderman, M.S., 2011, \aap, 525, 4
\bibitem[Ruderman \& Erd{\'e}lyi(2009)]{RudermanErdelyi2009}
Ruderman, M.~S., \& Erd{\'e}lyi, R., 2009, \ssr, 149, 199
\bibitem[Ruderman et~al.(2008)]{ruderman2008}
Ruderman, M.~S., Verth, G., \& Erd{\'e}lyi, R., 2008, \apj, 686, 694
\bibitem[Selwa et~al.(2010)]{selwa10}
Selwa, M., Murawski, K., Solanki, S.~K., \& Ofman, L., 2010, \aap, 512, A76
\bibitem[Ugarte-Urra et~al.(2002)]{ugarte2002}
Ugarte-Urra, I., Doyle, J. G., \& Madjarska, M. S., 2002, SOLMAG 2002.~ Proceedings of the Magnetic Coupling of the Solar Atmosphere Euroconference, 505, 595
\bibitem[Van Doorsselaere et~al.(2007)]{vandoorsselaere2007}
Van Doorsselaere, T., Nakariakov, V.~M., \& Verwichte, E., 2007, \aap, 473, 959
\bibitem[Van Doorsselaere et~al.(2008)]{vandoors08}
Van Doorsselaere, T., Ruderman, M.S., \& Robertson, D., 2008, \aap, 485, 849
\bibitem[Verth et~al.(2007)]{verth2007}
Verth, G., Van Doorsselaere, T., Erd{\'e}lyi, R., \& Goossens, M., 2007, \aap, 475, 341
\bibitem[Verth \& Erd{\'e}lyi(2008)]{VerthErdelyi2008}
Verth, G., \& Erd{\'e}lyi, R.\ 2008, \aap, 486, 1015
\bibitem[Verth et~al.(2008)]{VerthErdelyiJess2008}
Verth, G., Erd{\'e}lyi, R., \& Jess, D.~B., 2008, \apjl, 687, L45
\bibitem[Verwichte et~al.(2004)]{verwichte2004}
Verwichte, E., Nakariakov, V.~M., Ofman, L., \& Deluca, E.~E., 2004, \solphys, 223, 77
\bibitem[Wills-Davey \& Thompson(1999)]{wills-davey1999}
Wills-Davey, M.~J., \& Thompson, B.~J., 1999, \solphys, 190, 467

\end{document}